\documentclass{PoS}
\usepackage[square,sort,comma,numbers]{natbib}
\usepackage{aas_macros}

\title{The off-axis jet structure in Mrk~501 at mm-wavelengths}

\ShortTitle{The off-axis jet structure in Mrk~501 at mm-wavelengths}

\author{\speaker{Shoko Koyama}$^{a}$, Motoki Kino$^{b}$, Marcello Giroletti$^{c}$, Akihiro Doi$^{d}$, Hiroshi Nagai$^{e}$, Kazuhiro Hada$^{c,e}$, Kotaro Niinuma$^{f}$, Monica Orienti$^{c}$, Gabriele Giovannini$^{g,c}$, Eduardo Ros$^{a,h,i}$, Tuomas Savolainen$^{j,a}$, Miguel. \'{A}. P\'{e}rez-Torres$^{k,l}$, and Thomas P. Krichbaum$^{a}$\\%, and Marco Chiaberge$^{m,c}$\\
        \llap{$^{a}$} Max-Planck-Institute f\"{u}r Radioastronomie, Auf dem H\"{u}gel 69, D-53121 Bonn, Germany\\
        \llap{	$^{b}$} Korea Astronomy and Space Science Institute, 776 Daedeokdae-ro, Yuseong-gu, Daejeon, 305-348, 
	Republic of Korea\\
	\llap{$^{c}$} INAF Istituto di Radioastronomia, via Gobetti 101, I-40129 Bologna, Italy\\
	\llap{$^{d}$} Institute of Space and Astronautical Science, Japan Aerospace Exploration Agency, 3-1-1 Yoshinodai, Chuou-ku, Sagamihara, Kanagawa 252-5210, Japan\\
	\llap{$^{e}$} National Astronomical Observatory of Japan, 2-21-1 Osawa, Mitaka, Tokyo 181-8588, Japan\\
	\llap{$^{f}$} Graduate School of Science and Engineering, Yamaguchi University, Yamaguchi 753-8511, Japan\\
	\llap{$^{g}$} Dipartimento di Astronomia, Universit\`{a} di Bologna, via Ranzani 1, I-40127 Bologna, Italy\\
	\llap{$^{h}$} Observatori Astron\`{o}mic, Universitat de Val\`{e}ncia, E-46980 Paterna, Val\`{e}ncia, Spain\\
	\llap{$^{i}$} Departament d'Astronomia i Astrof\'{i}sica, Universitat de Val\`{e}ncia, E-46100 Burjassot, Val\`{e}ncia, Spain\\
	\llap{$^{j}$} Aalto University Mets\"{a}hovi Radio Observatory, FIN-02540 Kylm\"{a}l\"{a}, Finland\\
	\llap{$^{k}$} Instituto de Astrof\'{i}sica de Andaluc\'{i}a, Glorieta de las Astronom\'{i}a, s/n, E-18008 Granada, Spain\\
	 \llap{$^{l}$} Centro de Estudios de la F\'{i}sica del Cosmos de Arag\'{o}n, E-44001 Teruel, Spain\\
	E-mail: \email{skoyama@mpifr-bonn.mpg.de}\\}

\abstract{We present results from 43~GHz (VLBA, six epochs from 2012.2 to 2013.2) 
and 86~GHz (GMVA, one epoch in 2012.4) observations toward the basis of the jet in the TeV Blazar Mrk~501. 
The 43-GHz data analysis reveals a new feature located northeast of the radio core, 
with a flux density of several tens of mJy, perpendicularly to the jet axis. 
The 86-GHz image shows the jet feature located 0.75~mas southeast of the radio core, 
which is consistent with the previous result. 
The location of Gaussian model for 0.75~mas feature does not coincide with those for the jets in the 43-GHz image, 
however, a distribution of emission is found. 
We also discuss the spectral indices of the core, the northeast feature, 
and the jet feature between 43~GHz and 86~GHz, 
which show flat-to-steep, steep, and flat-to-invert, respectively.}

\FullConference{12th European VLBI Network Symposium and Users Meeting,\\
		7-10 October 2014\\
		Cagliari, Italy}

\begin{document}

\section{Introduction}
The innermost structure of relativistic jets in active galactic nuclei
are considered to be related to the properties of the regions
where the jets are formed, collimated, and accelerated.
Ultra-high resolution Very Long Baseline Interferometry (VLBI) observations
are the best tools to explore the parsec scale structure of jets.
Recently, 
the Event Horizon Telescope at 230~GHz (1.3~mm)
has revealed 
the off-axis jet structure within 0.3~mas from the core 
in two quasars
\citep{Lu:2012,Lu:2013}.
However, in BL Lac Objects,
such off-axis jet structure have not been studied.

The TeV blazar Mrk~501 is one of the best BL Lac Objects to resolve the innermost jet because of its proximity ($z=0.034$, $1~{\rm mas}=0.66$~pc) and brightness. 
The VLBI images show different jet position angles at different scales
\citep{Giroletti:2004}.
The previous Very Long Baseline Array (VLBA) 43~GHz and Global mm-VLBI Array (GMVA) 86~GHz 
images show clear limb-bright structure \citep{Giroletti:2008, Piner:2009}, 
while they do not show clear off-axis jet structure at the innermost part.

\section{Observations and Data Reduction}
We observed this target in six different epochs by the VLBA at 43~GHz and one epoch by the GMVA at 86~GHz,
with 8~IFs of 16~MHz bandwidth each.
%
%Left hand circular polarization (LHCP) was recorded at a bit-rate of 512~Mbps by Mark 5 disk systems.
%
The total on-source time for Mrk~501 was approximately one hour for the 43~GHz data,
and four hours for the 86~GHz data.
Phase and delay offset between different sub-bands were solved by using the calibrator 3C~345.
For the 43~GHz data,
after the standard initial calibrations in AIPS,
the final images were produced after iterations of CLEAN, phase and amplitude self-calibration processes in difmap.
For the 86~GHz data,
fringe fitting on Mrk~501 
were performed
with a small search window of a few tens of nsec and mHz in delay and rate respectively, 
and a low SNR threshold,
after applying the delay and rate solutions of 3C~345.
Fringe solutions of Mrk~501 were not derived among transatlantic baselines,
however, weak coherent phases were found.
Gaussian model fitting and phase self-calibration were done in difmap,
without amplitude self-calibration due to low SNR of the visibilities.

\section{Results}
\begin{figure*}
\centering
\includegraphics[width=11.5cm]{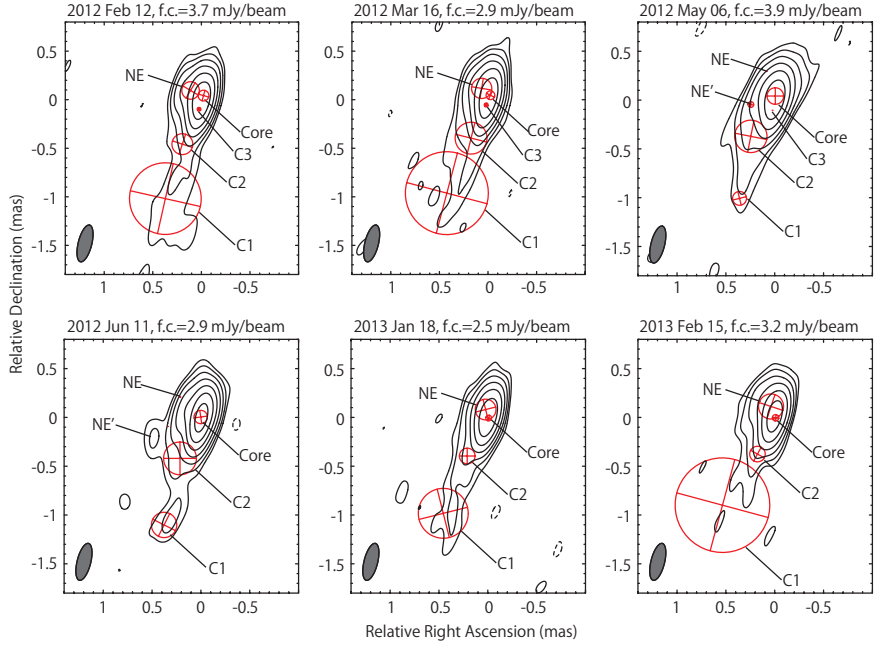}
\caption{43~GHz uniform weighted VLBA images with fitted circular Gaussian components.
The restoring beam is 0.39~mas $\times$ 0.14~mas in $PA~-14^{\circ}$, plotted in the bottom-left corner.
Observing date and the first contour (f.c.) are plotted above each map.
The first contour are all set to three times the rms noise level of each map, increasing by a factor of 2. }
\label{res01835}
\end{figure*}

In Figure~\ref{res01835}, 
we find a new component (labeled NE) located 
almost perpendicularly to the main jet axis (labeled C1, C2, C3).
The location (angular separation, position angle, and flux) of NE changes randomly.
We also confirmed the existence of NE
using the slice profiles of the CLEAN images
across the center of Gaussian models for core and NE. %(Figure~\ref{slice} {\it left}).
There is a clear emission at the northeast of the core for all epochs
with the peak flux of $36~{\rm mJy~beam^{-1}}$, 
the position of $0.18$~mas from the core,
and the FWHM of $0.24$~mas in average.
These results suggest that the limb-brightened structure of Mrk~501
apparently stars in the nearest vicinity of the core
due to the small viewing angle.

Figure~2 {\it -Left} shows the total-intensity image
obtained by the GMVA at 86~GHz
at a resolution of 205~$\mu$as $\times$ 37~$\mu$as ($PA~-5.11^{\circ}$)
with fitted Gaussian models.
We find a tentative jet knot located at $(r,~\theta)=(0.75~{\rm mas}, 159^{\circ})$
with a SNR of $\sim5\sigma$.
This knot could correspond to one detected by the GMVA at 86~GHz in \cite{Giroletti:2008},
located at $(r,~\theta)=(0.73~{\rm mas}, 172^{\circ})$.
The flux of the core is $197\pm59$~mJy and that of the tentative jet knot is $19\pm8$~mJy.

Since the 86-GHz data and the third-epoch VLBA data at 43~GHz were obtained within two weeks,
we compare these two images.
In the 43-GHz image,
there are no Gaussian models at $\sim0.75$~mas southeast of the core,
however,
a distribution of emission is found.
The spectral index of the integrated core fluxes over Gaussian models between 43~GHz and 86~GHz is $\alpha=-0.8\pm0.5$,
where we define the flux density at observing frequency $\nu$ as $S_{\nu}\propto\nu^{\alpha}$.
This flat-to-steep spectral tendency is consistent 
with the power law index of the core reported in \cite{Giroletti:2008} 
($\alpha\sim-0.5$ above 8~GHz).
Using the model-fit flux of the 43-GHz image and the $3\sigma$ upper limit in the 86-GHz image,
the spectral index of NE is estimated to be 
$\alpha\le-0.8$, which is steeper than that for the radio core.
We also estimate the spectral index at the southeast jet as $\alpha=-0.1\pm0.5$ 
by integrating the flux within the same region between the two images using the AIPS task IMSTAT.
%
%The integrated flux in the VLBA 43~GHz image is $8.4\pm0.8$~mJy,
%while that in the GMVA 86~GHz image is comparable, that is, $9.0\pm2.7$~mJy.
%
Due to the large spectral index error, this is not a stringent result.
However, despite of the steep spectrum of the core, 
the southeast jet spectrum is flat-to-inverted
and it is inconsistent with one at lower frequencies \citep[e.g.,][]{Hovatta:2014}.
This tendency is also seen in the spectral index map shown in Figure~2 {\it -Right}.

%\includegraphics[width=7cm]{501vlba_rad_20141002.eps}
%\caption{Angular separation from the core to each component. C1 (blank triangles), C2 (blank squares), C3 (filled triangles), NE (crosses), and NE$'$ (filled squares).}
%\label{vlbarad}
%\includegraphics[width=7cm]{501vlba_pa_20141002.eps}
%\caption{Position angle of each component relative to the core. C1 (blank triangles), C2 (blank squares), C3 (filled triangles), NE (crosses), and NE$'$ (filled squares).}
%\label{vlbapa}
%\includegraphics[width=7cm]{501vlba_flux_20141002.eps}
%\caption{Light curves of each component (C1, C2, C3, Core, NE, NE$'$), total CLEANed flux, and peak flux.}
%\label{vlbalight}
%\includegraphics[width=5cm]{GG072_BK172C_overlay.eps}
%\includegraphics[width=5cm]{BK172slices2.eps}

\begin{figure}
\begin{minipage}{0.5\hsize}
\centering
\includegraphics[width=4cm]{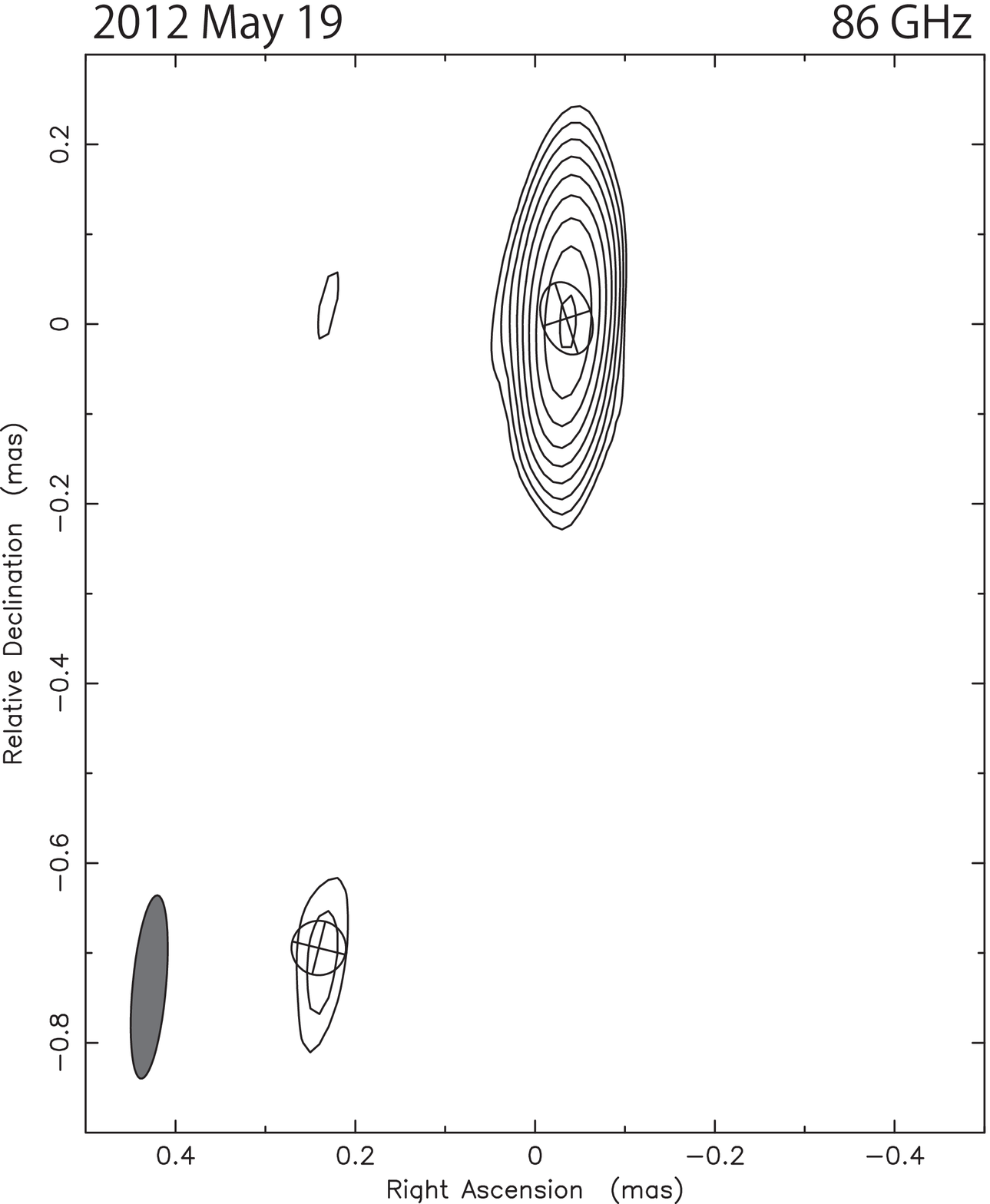}
\label{gg072}
\end{minipage}
\begin{minipage}{0.5\hsize}
\centering
\includegraphics[width=8cm]{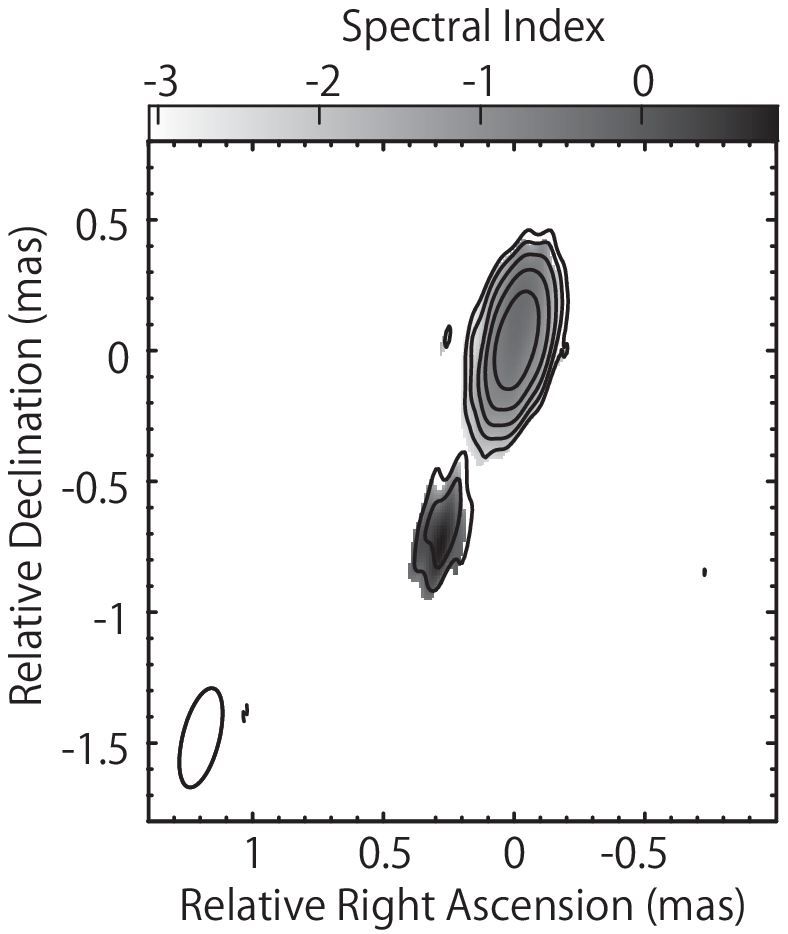}
\label{spix}
\end{minipage}
\caption{{\it Left}: GMVA image of Mrk~501 at 86 GHz with model fit components with the restoring beam of 205~$\mu$as $\times$ 37~$\mu$as in $PA~-5.11^{\circ}$. The peak brightness is 98~${\rm mJy~beam^{-1}}$, and the contour levels are drawn at ($-$1,~1,~1.141,~2,…)$\times5.8~{\rm mJy~beam^{-1}}$. The 1$\sigma$ noise level is $1.9~{\rm mJy~beam^{-1}}$.
{\it Right}: Spectral index map between the GMVA 86 GHz image and the third epoch VLBA 43 GHz image with total intensity contours of restored GMVA image.}
\end{figure}

\section{Summary and future prospects}
We have explored the parsec-scale regions in the jet of Mrk~501 using the VLBA at 43 GHz and the GMVA at 86 GHz.
We newly find the off-axis jet structure, NE, in the 43 GHz image,
and detect a 0.75~mas jet feature in the 86~GHz image.
The spectrum of the radio core is flat-to-steep and consistent with previous results, 
while the jet spectrum is flat-to-invert and inconsistent with one at lower frequencies.

This observation cannot confirm the spectral index of the jet feature in Mrk~501
due to the limited ($u,v$)-coverage provided by the array used.
To obtain more precise GMVA images,
a better ($u,v$)-coverage, 
achievable by including new stations (e.g., the Korean VLBI Network) in the array,
is indeed recommended.

\begin{small}
\paragraph*{Acknowledgments}
%This work made use of the Swinburne University of Technology software correlator \citep{Deller:2011}, developed as part of the Australian Major National Research Facilities Programme and operated under license.
This paper is partially based on observations carried out with the VLBA, the MPIfR 100 m Effelsberg Radio Telescope, the IRAM Plateau de Bure Millimetre Interferometer, the IRAM 30 m Millimeter Telescope, the Onsala 20 m Radio Telescope, and the Mets\"{a}hovi 14 m Radio Telescope.
%IRAM is supported by MPG (Germany), INSU/CNRS (France), and IGN (Spain). 
The National Radio Astronomy Observatory is a facility of the National Science Foundation operated under cooperative agreement by Associated Universities, Inc. 
The GMVA is operated by the MPIfR, IRAM, NRAO, OSO, and MRO. 
%We thank the staff of the participating observatories for their efficient and continuous support.
Part of this work was done with the contribution of the Italian Ministry of Foreign Affairs and University and Research for the collaboration project between Italy and Japan.
\end{small}

\bibliographystyle{JHEP}
\bibliography{/Users/kshoko/Documents/Astro}
\begin{footnotesize}
\providecommand{\href}[2]{#2}\begingroup\raggedright\endgroup

\end{footnotesize}

\end{document}